\documentclass[aps,prm,reprint,superscriptaddress]{revtex4-2}
\usepackage{graphicx}
\usepackage{dcolumn}
\usepackage{bm}

\newcommand{\pn}{$\mathrm{P{_{N_2}}}$}
\newcommand{\pa}{$\mathrm{P{_{Ar}}}$}
\newcommand{\p}{$\%$}

\newcommand{\Ts}{T$_\mathrm{s}$}

\newcommand{\rn}{$\mathrm{R{_{N_2}}}$}
\newcommand{\Hf}{$ \Delta $H$ _{f}^{0} $}
\newcommand{\Eg}{E$_\mathrm{g}$}

\begin{document}

\title{Synthesis and study of ScN thin films}
\author {Susmita Chowdhury}
\affiliation{Applied Science Department, Institute of Engineering and Technology, DAVV, Indore, 452017, India}
\author{Rachana Gupta}
\affiliation{Applied Science Department, Institute of Engineering and Technology, DAVV, Indore, 452017, India}
\author{Parasmani Rajput}
\affiliation{Beamline Development and Application Section, Physics Group, Bhabha Atomic Research Centre, Mumbai 400085, India}
\author{Akhil Tayal}
\affiliation{Deutsches Elektronen-Synchrotron DESY, Notkestrasse 85, D-22607 Hamburg, Germany}
\author{Dheemahi Rao}
\affiliation{Chemistry and Physics of Materials Unit, Jawaharlal Nehru Centre for Advanced Scientific Research, Bengaluru 560064,India}
\affiliation{International Centre for Materials Science, Jawaharlal Nehru Centre for Advanced Scientific Research,Bengaluru 560064, India}
\affiliation{School of Advanced Materials (SAMat), Jawaharlal Nehru Centre for Advanced Scientific Research, Bengaluru 560064,India}
\author{ Reddy Sekhar}
\affiliation{Surface and Nanoscience Division, Materials Science Group, Indira Gandhi Centre for Atomic Research, HBNI, Kalpakkam - 603102, India}
\author{Shashi Prakash}
\affiliation{Applied Science Department, Institute of Engineering and Technology, DAVV, Indore, 452017, India}
\author{Ramaseshan Rajagopalan}
\affiliation{Surface and Nanoscience Division, Materials Science Group, Indira Gandhi Centre for Atomic Research, HBNI, Kalpakkam - 603102, India}
\author{S. N. Jha}
\affiliation{Beamline Development and Application Section, Physics Group, Bhabha Atomic Research Centre, Mumbai 400085, India}
\author{Bivas Saha}
\affiliation{Chemistry and Physics of Materials Unit, Jawaharlal Nehru Centre for Advanced Scientific Research, Bengaluru 560064,India}
\affiliation{International Centre for Materials Science, Jawaharlal Nehru Centre for Advanced Scientific Research,Bengaluru 560064, India}
\affiliation{School of Advanced Materials (SAMat), Jawaharlal Nehru Centre for Advanced Scientific Research, Bengaluru 560064,India}
\author{Mukul Gupta}
\email[Corresponding author:]{mgupta@csr.res.in}
\affiliation{UGC-DAE Consortium for Scientific Research, University Campus, Khandwa Road, Indore-452 001,India}
\date{\today}

\begin{abstract}

To contemplate an alternative approach for the minimization of diffusion at high temperature depositions, present findings impart viability of room-temperature deposited reactively sputtered ScN thin film samples. The adopted room temperature route endows precise control over the {\rn} flow for a methodical structural phase evolution from Sc$ \rightarrow $ScN and probe the correlated physical aspects of the highly textured ScN samples. In the nitrided regime i.e. at {\rn} = 2.5~-~100{\p} flow, incorporation of unintentional oxygen defects were evidenced from surface sensitive soft x-ray absorption spectroscopy study, though less compared to their metal ({\rn} = 0{\p}) and interstitial ({\rn} = 1.6{\p}) counterparts, due to higher Gibb's free energy for Sc-O-N formation with no trace of ligand field splitting around the O K-edge spectra. To eradicate the sceptism of appearance of N K-edge (401.6\,eV) and Sc L-edge (402.2\,eV) absorption spectra adjacent to each other, the nascent Sc K-edge study has been adopted for the first time to validate complementary insight on the metrical parameters of the Sc-N system taken into consideration. Optical bandgaps of the polycrystalline ScN thin film samples were found to vary between 2.25~-~2.62\,eV as obtained from the UV-Vis spectroscopy, whereas, the nano-indentation hardness and modulus of the as-deposited samples lie between 15~-~34\,GPa and 152~-~476\,GPa, respectively following a linearly increasing trend of resistance to plastic deformations. Besides, contrary to other early 3d transition metal nitrides (TiN, VN, CrN), a comprehensive comparison of noticeably large homogeneity range in Sc-N has been outlined to apprehend the minuscule lattice expansion over the large {\rn} realm.   

\end{abstract}

\maketitle
\section{Introduction}
\label{INTRODUCTION}

Early 3d transition metal nitrides (TMNs) e.g. ScN, TiN, VN and CrN even though crystallizes in cubic rocksalt-type B1 structure, but distinct band structure manifests heterogeneous electrical conductivity within the family of early TMNs~\cite{saha2018rocksalt,patsalas2018conductive}. Among them, ScN as a semiconductor sought a profound research attention in recent times, whereas, the rest of the early 3d TMNs exhibit metallic nature and are substantially well explored since  decades~\cite{patsalas2018conductive,biswas2019development}. Apart from ScN been a pre-eminent refractory compound (melting point exceeding $ \approx $2873\,K, corossion resistant, high hardness of $ \approx $21\,GPa)~\cite{biswas2019development} exhibiting high thermoelectric \textit{figure-of-merit} (0.3 at
800\,K)~\cite{burmistrova2013thermoelectric} and assist as a template for the growth of low dislocation density GaN~\cite{scholz2012semipolar}, ScN further possess immense functionalities in conjuction with other TMNs viz. Sc$ _{x} $Ga$ _{1-x} $N as light emitting diodes~\cite{little2001band}, Al$ _{1-x} $Sc$ _{x} $N as MEMS magnetoelectric sensors~\cite{su2020alscn,tasnadi2010origin}, epitaxial (Zr,W)N/ScN~-~metal/semiconductor superlattices as thermionic energy conversion devices~\cite{rawat2009thermal} etc. Besides these intriguing aspects, the lowest enthalpy of formation ({\Hf} = -19.79\,eV)~\cite{belosludtsev2018correlation} Sc-O of Sc compared to other TMNs is the principal intricacy for synthesis of pure ScN, resultant being an unintentional n-type degenerate semiconductor~\cite{saha2017compensation,eklund2016transition}. 

Hence for application based perspectives, to modulate the band structure engineering for superior device performances, so far, most of the studies adopted high vacuum ($ \leq $10$ ^{-8} $\,Torr) depositions to ensure low defect concentrations with atomically smooth epitaxial growth of ScN thin film samples on variety of single crystal substrates (MgO, Al$ _{2} $O$ _{3} $, GaN, SiC etc.)~\cite{more2020correlation,rao2020high,casamento2019molecular,le2018effect}, and few of them aided with process parameters are tabulated in Table~\ref{techniques}. As can be seen from Table~\ref{techniques}, conventional use of high substrate temperature ({\Ts} $ \geq $ 823\,K) has been an integral part during the synthesis of ScN thin film samples, possibly due to higher adatom mobility promoting enhanced crystalline defect free ScN growth~\cite{rao2020effects}. In addition, intensive research attention have also been dedicated to get an insight on explicit defect contributions and microstructural growth behavior (e.g. dislocations, twin domains etc.) of ScN samples and when fabricated with other metals and/or TMNs as in metal/semiconductor superlattices, multilayers etc~\cite{saha2018rocksalt,burmistrova2013thermoelectric,kumagai2018point,acharya2021twinned}. 

In terms of defects, even though it is well known that during the growth of ScN itself, finite incorporation of substitutional (O$ _\mathrm{N} $) and/or interstitial oxygen (O$ _\mathrm{i} $) is inherent regardless of {\Ts}~\cite{le2018effect,kumagai2018point,kumar2021clustering}, but combined study of first principles density functional theory (DFT) with site occupancy disorder technique reveals that the electronic band structure of ScN remains unaltered despite of a shift of the Fermi energy level to the bottom of the conduction band~\cite{burmistrova2013thermoelectric}. Nonetheless, only recently, the primary contribution of oxygen incorporation has also been attributed to the surface oxidation~\cite{more2020correlation,rao2020effects}. Howbeit, nitrogen vacancies (V$ _\mathrm{N} $) are known to form a defect energy level at $ \approx $1.26\,eV above the valence band maxima at $ \Gamma $ point of the Brillouin zone~\cite{haseman2020cathodoluminescence}. In this context, it is to be mentioned here that significance of high {\Ts} depositions in supression of defects were found to be conflicting in literature~\cite{more2020correlation,rao2020effects,moram2008effect,ohgaki2013electrical}, yet have not been highlighted so far. 

Furthermore, as regards to technological viability in electronics viz. CMOS integrated circuits, plastic substrates etc, high {\Ts} synthesis is highly undesirable~\cite{tsai2021room}. Moreover, the extent of diffusion across the metal-semiconductor superlattice and/or multilayer interfaces will be comparatively high at a high {\Ts} regime~\cite{eklund2016transition}, which could limit the device performances in practical applications. Additionally, interdiffusion across film-substrate interfaces are also pronounced at high {\Ts} depositions~\cite{haseman2020cathodoluminescence,cetnar2018electronic}. In view of this, contrary to high {\Ts} depositions, we adopted a room temperature deposited reactive magnetron sputtering technique for the synthesis of ScN thin film samples, as ScN favors thermodynamical growth conditions even at 298\,K ({\Hf} = -3.29\,eV). Such temperature regime also paves the way for precise control over variation of relative N$ _{2} $ partial pressures (\rn) to probe the structural phase evolution from hexagonal close packed (hcp) Sc to rocksalt-type face centered cubic (fcc) ScN, which is still ambiguous. 

In order to probe the electronic structure of ScN, so far, usually x-ray photoelectron spectroscopy (XPS) or soft x-ray absorption spectroscopy (SXAS) at N K-edge (401.6\,eV) and Sc L-edge (402.2\,eV) were considered~\cite{haseman2020cathodoluminescence,more2019electrical,nayak2019rigid,biswas2020interfacial}. But, since the two absorption edges appear very close to each other and moreover, both XPS and SXAS are known to be surface sensitive techniques~\cite{henderson2014x}, an alternative powerful technique such as x-ray absorption fine structure (XAFS) can provide better insight on the metrical parameters at atomic scale level. With this motif, for the first time, XAFS was implemented on the Sc-N system complementary to SXAS to probe the K-edge of Sc in ScN. In spite of recent surge in investigation of various physical properties, further realization of variation of N were systematically demonstrated in terms of structural, electronic, optical and mechanical responses of room temperature deposited ScN thin film samples which are still missing in literature.

\begin{table*} \caption{\label{techniques} Growth techniques of ScN thin film samples deposited using various substrate temperature (\Ts) and deposition power (P) on different substrates with corresponding lattice parameter (LP) and direct optical bandgap ({\Eg}) values. Here, dcMS = Direct current magnetron sputtering and MBE = Molecular beam epitaxy, RMS = Reactive magnetron sputtering, 300$ ^{\dagger}$ = amorphous at 300\,K, GGA-PBE = Perdew-Burke-Ernzerhof GGA exchange correlation functional, HSE06 = Heyd-Scuseria-Ernzerhof Hybrid functional, FLAWP = Full-potential linearized augmented plane wave method, FLAWP-GGA = Full-potential linearized augmented plane wave method with generalized gradient approximation, LDA = Local density approximation, GGA+U = Generalized gradient approximation with Hubbard \textit{U} correction.}
	
	\begin{tabular} {llllll} \hline
		\textbf{Exp.} & \textbf{Substrate} & \textbf{Process}  & \textbf{LP} & \textbf{{\Eg}} & \textbf{Ref.} \\
		\textbf{Tech.} & & \textbf{Parameters} & \textbf{(\AA)} & \textbf{(eV)} &\\
		\hline 
		dcMS & MgO (001) & {\Ts} = 1103\,K, & 4.50 & - &\cite{burmistrova2013thermoelectric}\\
		& \& Si (001) & P =150\,W &  &  &\\
		dcMS& c-plane Al$ _{2} $O$ _{3} $, & {\Ts} = 973~-~1223\,K,  & 4.504 - 4.512 & - & \cite{le2018effect}\\
		& MgO (111) \& & P = 125\,W &  &  & \\
		& r-plane Al$ _{2} $O$ _{3} $ & & & &\\
		dcMS & MgO (001) & {\Ts} = 1123 \& 1223\,K, & 4.50 & 2.18 - 2.7 & \cite{deng2015optical}\\
		& & P = 60~-~300\,W &  &  & \\
		dcMS & MgO (001) & {\Ts} = 973\,K & 4.573 & 2.59 & \cite{more2019electrical}\\
		MBE & GaN (0001), & {\Ts} = 1023\,K, & 4.497 & 2.1 & \cite{casamento2019molecular}\\
		& SiC (0001) \& & P = 200\,W &  &  & \\
		& AlN (0001) & & & &\\
		RMS & MgO (001) & {\Ts} = 823\,K, & 4.52 - 4.54 & 2.1 & \cite{cetnar2018electronic}\\
		& & P = 25 - 127\,W & & & \\
		dcMS & MgO (001) & {\Ts} = 873 - 1073\,K, & 4.50 & 2.19 - 2.23 & \cite{rao2020effects}\\
		&  & P = 125\,W & & & \\
		dcMS & c-plane Al$ _{2} $O$ _{3} $ & {\Ts} = 300$ ^{\dagger}$ - 1023\,K, & $ \approx $4.47 - 4.52, & 2.2 - 3.1 & \cite{rao2020effects}\\
		& \& Si & P = 40\,W & & & \\
		MBE & MgO (001), & {\Ts} = 1073\,K, & - & 2.15 & \cite{al2004surface}\\
		\textbf{dcMS} & \textbf{Quartz \& } & \textbf {\Ts~=~300\,K,} & \textbf{4.49 - 4.567} & $ \textbf{2.25 - 2.62} $ & \textbf{this}  \\
		&  \textbf{Si (100)} & \textbf{P = 100\,W} &  &  & \textbf{work} \\
		\hline
		\textbf{Theoretical} &  & & & & \\\hline
		GGA-PBE & - & - & 4.519  & 2.02 & \cite{deng2015optical}\\
		\& HSE06 &  &  & 4.499 &- & \\
		FLAPW \& & - & - & 4.42 & - & \cite{stampfl2001electronic}\\
		FLAPW-GGA & - & - & 4.50 & -& \\
		LDA & - & - & 4.47 & -& \cite{gall2001vibrational}\\
		GGA+\textit{U} & - & - & 4.52 & 1.86 & \cite{saha2010electronic}\\\hline
		
		\end{tabular}
\end{table*}

\section{Experimental}

Metallic Sc and a series of ScN thin film samples were deposited on amorphous quartz and single crystal Si (100) substrates at various {\rn} [= {\pn}/({\pn} $ + $ {\pa}), where {\pn} and {\pa} are nitrogen and argon partial pressures, respectively] flow = 1.6, 2.5, 5, 10, 25, 50 and 100{\p} in closed intervals using a direct current magnetron sputtering (dcMS) at ambient temperature ($ \approx $300\,K). For thin film deposition, a Sc (99.95{\p} pure) 3-inch target was sputtered in the presence of 5N purity Ar and/or N$ _{2} $ gas flows. Prior to the deposition, the substrates were cleaned in an ultrasonic bath of acetone followed by methanol wiping with dry air blown and were loaded into the chamber. Subsequently, the sample holder was baked for 1 hour at 573\,K and then cool down to room temperature to achieve a base-pressure of about 1$ \times $10$ ^{-7} $\,Torr or lower. During deposition, the working pressure was $ \approx $3$ \times $10$ ^{-3} $\,Torr and the substrate holder rotation was kept fixed at 60\,rpm to get better uniformity of the samples. 

For thickness calibration of the samples, x-ray reflectivity (XRR) measurements were performed using Cu-K$\alpha$ x-rays on a Bruker D8 Discover system. Once the deposition rate was obtained from the fitting of the XRR data (not shown), typically 200\,nm thick samples were prepared following the similar deposition procedure. The structural characterization of samples were carried out using x-ray diffraction (XRD) using a Bruker D8 Advance
XRD system based on $\theta$-2$\theta$ Bragg-Brentano geometry with Cu-K$\alpha$ (1.54\,{\AA}) x-rays and detected using a fast 1D detector (Bruker LynxEye).
To probe the local electronic structure, surface sensitive SXAS measurements were performed at N K-edge and Sc L$ _\mathrm{(III,II)}$-edges in total electron yield (TEY) mode at BL-01 beamline of Indus-2 synchrotron radiation source~\cite{2014_AIP_BL01} at RRCAT, Indore,
India. Complementary to SXAS, to get an elementary insight probing the deep core level in atomic scale regime, x-ray absorption fine structure (XAFS) measurements were performed in fluorescence mode at BL-09 beamline at RRCAT, Indore, India and also at P64 beamline of PETRA-III, DESY, Germany~\cite{caliebe2019high}. XAFS data taken at Sc K-edge from both beamlines were found to be similar and XANES data taken at BL-09 and EXAFS data taken at P64 has been included. The obtained data was processed in Athena software~\cite{ravel2005athena} with pre and post-edge normalization~\cite{teo1981extended} and fitting of the Fourier Transform (FT) spectra were performed using a software code developed by Conradson et al~\cite{conradson2013possible}. The fitted \textit{R} range was taken from 0 to 10\,{\AA}, while the used \textit{k}-range was 3 to 8\,\AA$ ^{-1} $.

The optical absorption spectra of the ScN thin film samples were recorded by Perkin Elmer, Lambda-750 UV–Visible spectrophotometer with double beam monochromator in the wavelength range of 250~-~1000\,nm at room temperature. The reflectance of the recorded data were converted to absorption spectra using Kubelka-Munk radiative transfer model, which is associated with the absorption coefficient ($ \alpha $)~\cite{murphy2007band} of the ScN thin film samples. To measure the hardness and elastic modulus of the samples, nanoindentation tests (Anton Paar, Switzerland) were performed using Berkovich diamond indenter tip with standard loading and unloading procedure based on Oliver and Pharr model~\cite{ramaseshan2016preferentially}. In order to suppress the substrate effects, the measurements were performed on one/tenth of the total sample thickness~\cite{panda2019effects}.  

\section{Results}\label{RESULTS}

\subsection{X-Ray Diffraction}\label{XRD}

\begin{figure}\centering
	\includegraphics
	[width=0.5\textwidth] {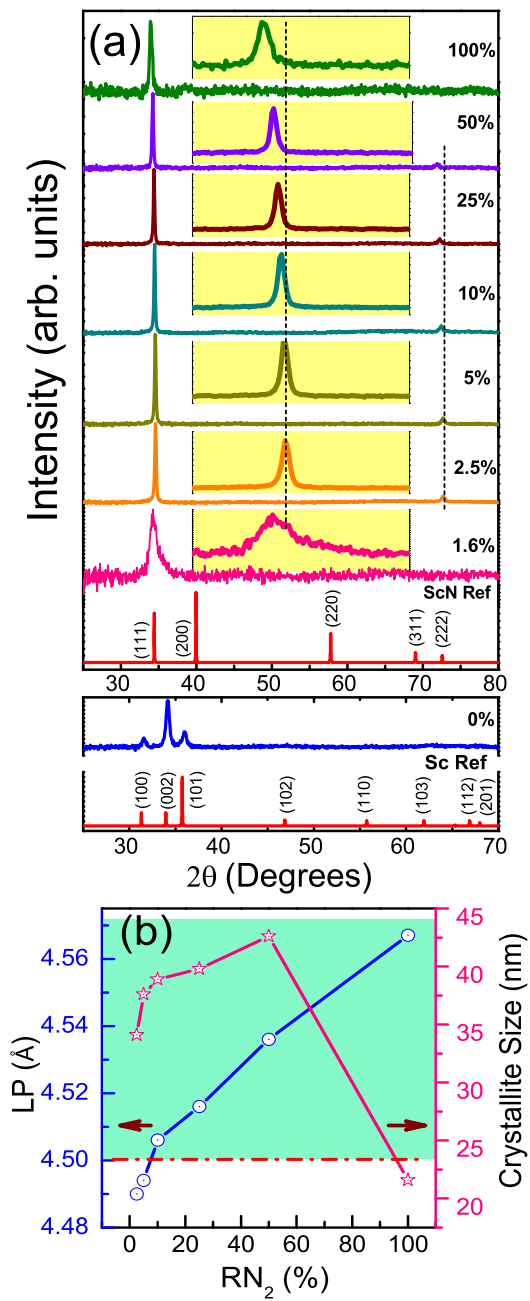}
	\vspace{-45mm}
	\caption{XRD pattern(a) and obtained variations in the lattice parameters and crytallite size (b) as a function of {\rn} of pure Sc and ScN samples deposited at various \rn = 1.6, 2.5, 5, 10, 25, 50 and 100{\p}.}
	\label{xrd}
\end{figure}

To take into account the phase formation of as-deposited samples, Figure~\ref{xrd}(a) illustrates the XRD data of Sc and ScN thin film samples and are compared with bulk references~\cite{deng2015optical,spedding1956crystal}, whereas, Figure~\ref{xrd}(b) demonstrates the obtained variations in the lattice parameters (LP) and crystallite size as a function of {\rn}. In addition, the highlighted region (in cyan) of Figure~\ref{xrd}(b) depicts the experimentally obtained LP of ScN thin film samples in literature and the red dotted line is a guide to eye for the theoretical predicted value of bulk ScN~\cite{deng2015optical}. As can be seen from Figure~\ref{xrd}(a), the occurence of three prominent peaks for Sc thin film sample can be assigned to (100), (002) and (101) reflection planes of hcp Sc, whereas for ScN samples, three different growth stages can be witnessed with variation in {\rn} flow namely, (i) interstitial incorporation of N atoms within hcp Sc, (ii) formation of NaCl rocksalt type fcc-ScN, and (iii) gradual expansion of ScN lattice due to incorporation of N atoms in fcc-ScN. 

Here, it is to be mentioned that an earlier report on deposition of polycrystalline ScN thin film samples on quartz substrates at {\Ts} = 300\,K using rf magnetron sputtering have reported amorphous growth and later on tuning of {\Ts} to high temperature resulted in preferential grain growth either along (111) or (200) plane, albeit the XRD data for the as-deposited samples were not presented therein~\cite{bai2001structure}. However, in the present work, at a very initial stage of {\rn}~=~1.6{\p}, the N atoms occupy the interstitial sites of hcp Sc manifesting an asymmetry and broadening in the reflection peak which suggests phase co-existance of hcp Sc and fcc ScN at certain phase fractions (as could also be evidenced from our XAFS data) and an enhancement in the crystalline disorder. Later on, when {\rn} was increased to 2.5{\p}, the sample exhibited a highly textured orientation with (111) and (222) reflection planes, resembling a rocksalt NaCl type structure even though the LP was 4.49\,{\AA}, slightly less than the theoretically predicted value of 4.501\,{\AA}~\cite{deng2015optical}. With further increase in {\rn} flow (from 5 to 50{\p}), the texturing of the samples remain unaltered, but due to gradual incorporation of N atoms within the crystal lattice, it starts to expand as shown in Figure~\ref{xrd}(a) by the gradual peak shifts of both (111) (magnified view shown in the highlighted inset) and (222) reflection peaks towards the lower diffraction angle (shown by dotted lines) accompanied with an increase in the crystallite size (Figure~\ref{xrd}(b)). 

Such unidirectional grain growth can be attributed to the kinetics driven mechanism at ambient temperature deposition ($ \approx $300\,K), due to trapping of the less mobile adatoms in the highest surface energy site i.e along the (111) reflection plane of ScN~\cite{gall1998microstructure}. Besides, further increase in {\rn} at 100{\p} flow causes lattice expansion in expense of reduced peak intensity of (111) reflection plane due to possible oversaturation of N, resulting in broadening of the (111) peak (as can be seen in highlighted region of Figure~\ref{xrd}(a)) with reduced crystallite size (Figure~\ref{xrd}(b)), and further absence of (222) grain growth suggests shattering of the long range periodicity. In view of growth evolution of Sc$\rightarrow$ScN at various {\rn} flow with electronic properties, SXAS measurements were performed and are discussed in section~\textbf{\ref{SXAS}}. 

\subsection{Soft X-Ray Absorption Spectroscopy}\label{SXAS}

\begin{figure*}\centering
	\includegraphics
	[width=1\textwidth] {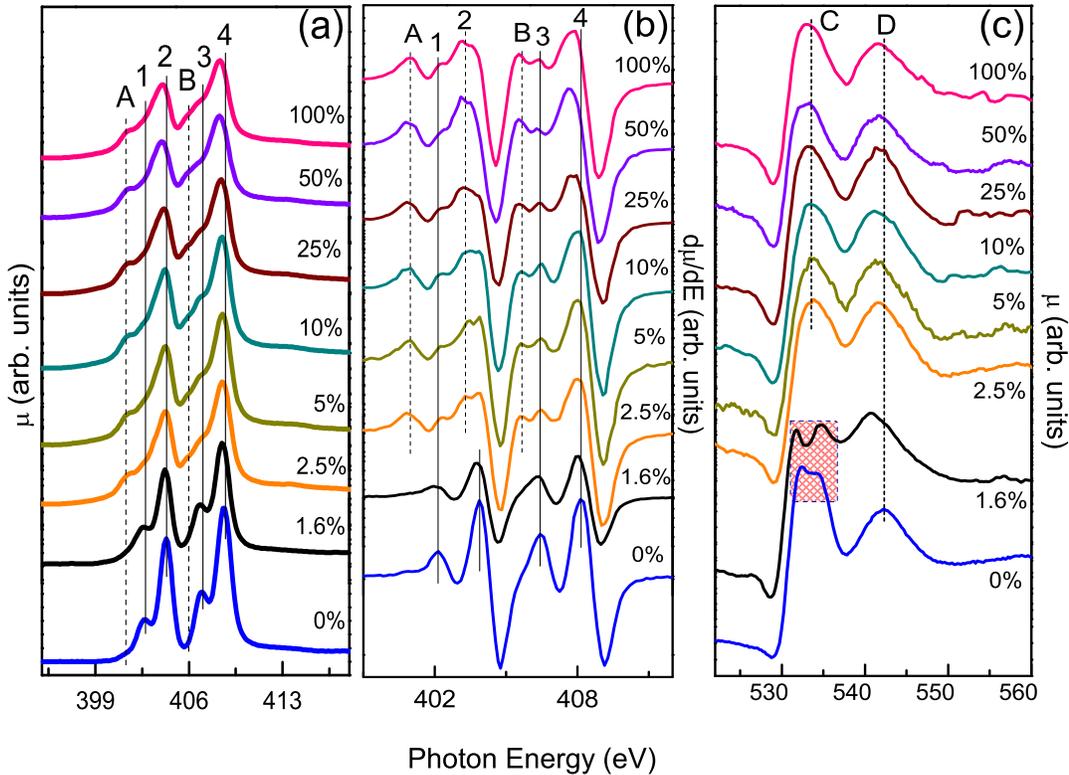}
	\vspace{-30mm}
	\caption{Normalized SXAS spectra of Sc L$ _\mathrm{III} $, L$ _\mathrm{II} $ and N K-edge (a), first order derivative of absorption spectra with respect to the photon energy (b) and O K-edge (c) of pure Sc and ScN samples deposited at various \rn = 1.6, 2.5, 5, 10, 25, 50 and 100{\p}.}
	\label{xas}
\end{figure*}

To get an insight on the local electronic structure of Sc and ScN thin film samples with variation in {\rn} flow, SXAS spectra of Sc L-edge and N K-edge were recorded and are shown in Figure~\ref{xas}(a), whereas, the first order derivative of the absorption spectra with respect to the photon energy is described in Figure~\ref{xas}(b). Additionally, to probe the oxidation effect, Figure~\ref{xas}(c) demonstrates the O K-edge XANES spectra of the samples. Here, it is to be mentioned that for a pure Sc sample, two absorption edges namely L$ _\mathrm{III} $ and L$ _\mathrm{II} $ are expected due to spin-orbit splitting of the Sc 2p orbital into 2p3/2 and 2p1/2 states in the absence of any ligand (C, N, O etc.) around the vicinity of Sc, a consequence of the transition of core electron from Sc 2p3/2$
\rightarrow $Sc 3d (L$ _\mathrm{III} $) and Sc 2p1/2$
\rightarrow $Sc 3d (L$ _\mathrm{II} $) states~\cite{chen1997nexafs}. However, in the present case, the prominent features of Sc sample in Figure~\ref{xas}(a) and \ref{xas}(b) marked as `1', `2', `3' and `4' can be assigned to L$ _\mathrm{III} $ (t$ _{2g} $), L$ _\mathrm{III} $ (e$ _{g} $), L$ _\mathrm{II} $ (t$ _{2g} $) and L$ _\mathrm{II} $ (e$ _{g} $) edges respectively due to the presence of a finite amount of unintentional O ligand field, which led to hybridization of Sc 3d-O 2p orbitals resulting in further splitting of each L$ _\mathrm{III} $ and L$ _\mathrm{II} $ features into t$ _\mathrm{2g} $ and e$ _\mathrm{g} $. 

Subsequently, the effect of surface oxidation/oxidation during the deposition itself for Sc sample is even pronounced from the O K-edge, where the doublet appearing at around 532.4 and 534.4\,eV (shown by the pink highlight) can be inferred to t$ _\mathrm{2g} $ (O 2p$\pi$+Sc3d) and e$ _\mathrm{g} $ (O 2p$\sigma$+Sc3d) states due to the possible octahedral ligand field splitting (10Dq), wheras the broad feature `D' arises due to hybridization of the O 2p with 4sp states of Sc~\cite{chen1997nexafs,de19902}. Similarly, for {\rn}~=~1.6{\p} sample, both Sc L-edge and O K-edge features mimic the same trend as metallic Sc thin film sample, but an increase in 10Dq $ \approx $ 2.9($ \pm $0.3)\,eV can be observed, which might be due to the shrink in volume of interstitial ScN during the process of structural transformation from Sc$ \rightarrow $ScN. Hence, O K-edge spectra confirms the incorporation of bonded O on the sample surface which can be present either in the form of Sc$ _\mathrm{x} $O$ _\mathrm{y} $ (for Sc)/ScO$ _\mathrm{x} $N$ _\mathrm{y} $ (for 1.6{\p} ScN), but, certainly not Sc$ _{2} $O$ _{3} $ (10Dq$ _\mathrm{[Sc-O]} $ = 3.3\,eV), as both samples were of metallic grey in color (as opposed to transparent Sc$ _{2} $O$ _{3} $).

With further increase in {\rn} from 2.5~-~100{\p}, as can be seen from Figure~\ref{xas}(a) and \ref{xas}(b), two new pronounced features labelled as `A' and `B' arises which can be ascribed to ligand field splitting in the presence of N octahedral environment and noticeable reduction in the intensity of features `1' and `3' can be detected which can be better discerned from the O K-edge spectra. Here, instead of a doublet, a new feature `C' can be noted. The appearance of similar experimental spectra have also been reported by Kumar et al. in the O K-edge of TiN ({\Ts} = 1023\,K) and through a combined study of DFT and ab-initio full potential multiscattering (FMS) theory, they concluded that such feature arises due to the presence of substitutional (O$ _\mathrm{N} $) and interstitial (O$ _\mathrm{i} $) oxygen in a defect complex state (4O$ _\mathrm{N} $+O$ _\mathrm{i} $)~\cite{kumar2021clustering}. Hence, the contribution of defects in early TMNs are comparable for both ambient and high {\Ts} depositions. As mentioned earlier, since the N K-edge and Sc L-edge appears very close to each other, Nayak et al. have performed theoretical simulations based on the FMS theory and reported a value of 10Dq = 2.1\,eV~\cite{nayak2019rigid}, which is in well agreement with the experimentally observed value of $ \approx $ 2.3($ \pm $0.3)\,eV obtained in the present work. Even though, it appears from Figure~\ref{xas}(a) that the energy features of `2' and `4' remain almost unaltered with an increase in {\rn} flow (2.5{\p} and above), but first order derivatives of the absorption spectra divulged a diverse profile where a new feature (around `2' and less pronounced for feature `4') towards the lower energy side for these samples is clearly visible as evidenced from Figure~\ref{xas}(b), which can be attributed as L$ _\mathrm{III} $ (e$ _{g} $) and L$ _\mathrm{II} $ (e$ _{g} $) corresponding to Sc-N bonds. Considering the new features of ScN, the spin-orbit splitting comes out to be 4.8($ \pm $0.3)\,eV, well in agreement as reported for ScN~\cite{biswas2020interfacial}.

\subsection{X-Ray Absorption Fine Structure}\label{EXAFS}

\begin{figure*}\center
	\includegraphics
	[width=0.75\textwidth] {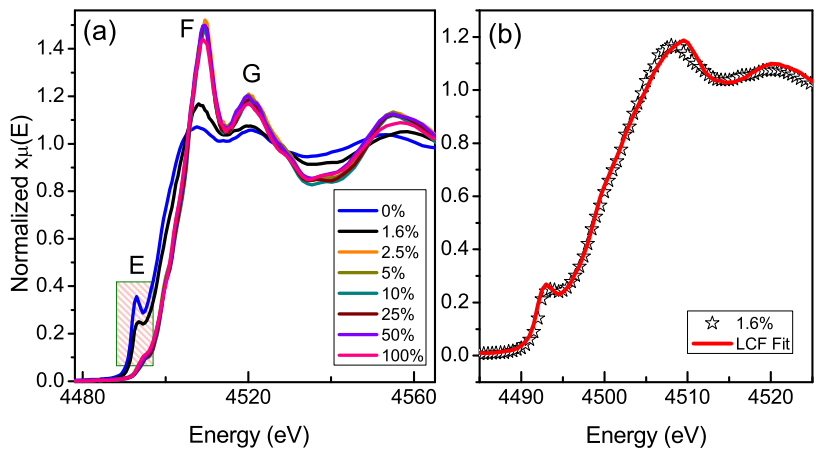}
	\vspace{-15mm}
	\caption{Normalized Sc K-edge XAFS spectra of metallic Sc and ScN thin film samples deposited at various \rn = 1.6, 2.5, 5, 10, 25, 50 and 100{\p} (a) and linear combination fit (LCF) of 1.6{\p} sample.}
	\label{exafsnorm}
\end{figure*}

So as to achieve complementary information about the electronic structure, Figure~\ref{exafsnorm}(a) depicts the normalized Sc K-edge XANES spectra of Sc and ScN samples deposited at various {\rn} flow and Figure~\ref{exafsnorm}(b) shows the linear combination fit (LCF) of 1.6{\p} sample. The spectra of Sc and 1.6{\p} ScN sample shows distinct features in comparison to samples deposited at relatively high {\rn} flow. In the pre-edge region, an intense feature `E' can be seen for Sc sample and with {\rn} flow at 1.6{\p}, it gets feeble. Such intense pre-edge feature has previously been reported for metallic hcp Ti (for both foil and film) with hexagonal symmetry and has been attributed to 1s $ \rightarrow $ 3d electric quadrupole transition ($ \Delta $\textit{l} = $ \pm $2)~\cite{yong2010ti,mardare2012meyer}. It is conventional that pre-edge features differ due to different geometrical parameters such as, inversion symmetry, co-ordinations and bonding configurations (e.g. bond length, bond angles) etc~\cite{jiang2007determination}. For 1.6{\p} sample, the best fit was obtained considering the phase co-existance of both Sc (71.9($ \pm $0.9)\,{\p}) and ScN (28.1($ \pm $0.9)\,{\p}) phases.

On the contrary, with further incorporation of N, the intensity gradually reduces from {\rn} = 1.6{\p} to 2.5{\p}, and are alike for rest of the samples. Since, it is well established that Sc is bonded with nearest neighbor N atoms in an octahedral co-ordination sphere with inversion symmetry~\cite{biswas2020interfacial}, the emergence of weak pre-edge feature can be ascribed to 1s core electron transition to 3d states of the absorber having a partial contribution from the 2p orbitals of N under the allowed electric dipole transition scheme ($ \Delta $\textit{l} = $ \pm $1), like in TiN~\cite{tuilier2008electronic}. 

In addition, nitridation of the samples consequences in continuous shift of the absorption edge (E$ _{0} $ = taken at 50{\p} of the absorption spectra)~\cite{kumar2021study} towards the higher energy side at E$ _{0} $ = 4496($ \pm $0.3) (Sc), 4497.4($ \pm $0.3) ({\rn} = 1.6{\p}) and 4500.3($ \pm $0.3) ({\rn} = 2.5{\p})\,eV which can be interpreted in terms of higher core-hole screening due to increase in valence states of Sc from Sc $\rightarrow$ ScN. However, above the absorption edge in the XANES region, the feature `F' and `G' can be assigned to 1s $ \rightarrow $ 4p electric dipole allowed transitions of a core electron as evidenced for other transition metal compounds e.g. TiC~\cite{moisy1988application}. A diverse trend of feature `F' can be attributed to the different stacking sequence of Sc (ABAB for hcp) and ScN (ABCABC for fcc) samples, where higher fcc phase fractions result in sharp intense peak (characteristic features of TMNs) in comparison to diffuse kind of feature for Sc thin film samples~\cite{longo2014crossing,tayal2020structural}, due to a possible intermixing between 3d quadrupole and 4p dipole states~\cite{henderson2014x}. It is worth mentioning here that in the present study, Sc K-edge of pure Sc thin film sample does not replicate the XAFS spectra of Sc$ _{2} $O$ _{3} $ studied by Chass\'{e} et al~\cite{chasse2018influence}, consistent with our XRD data analysis. Since, fluorescence detected XAFS is known to be a bulk sensitive technique with penetration depth ranging in micrometers~\cite{mardare2012meyer} compared to surface sensitive SXAS where the depth scale ranges only in nanometer scale ($ \approx $10\,nm)~\cite{henderson2014x}, the averaged out bulk information from the XAFS data further confirms the presence of higher surface oxidation in the samples as was evidenced from the Sc L-edge and O K-edge SXAS spectra.

\begin{figure*}\center
	\includegraphics
	[width=1\textwidth] {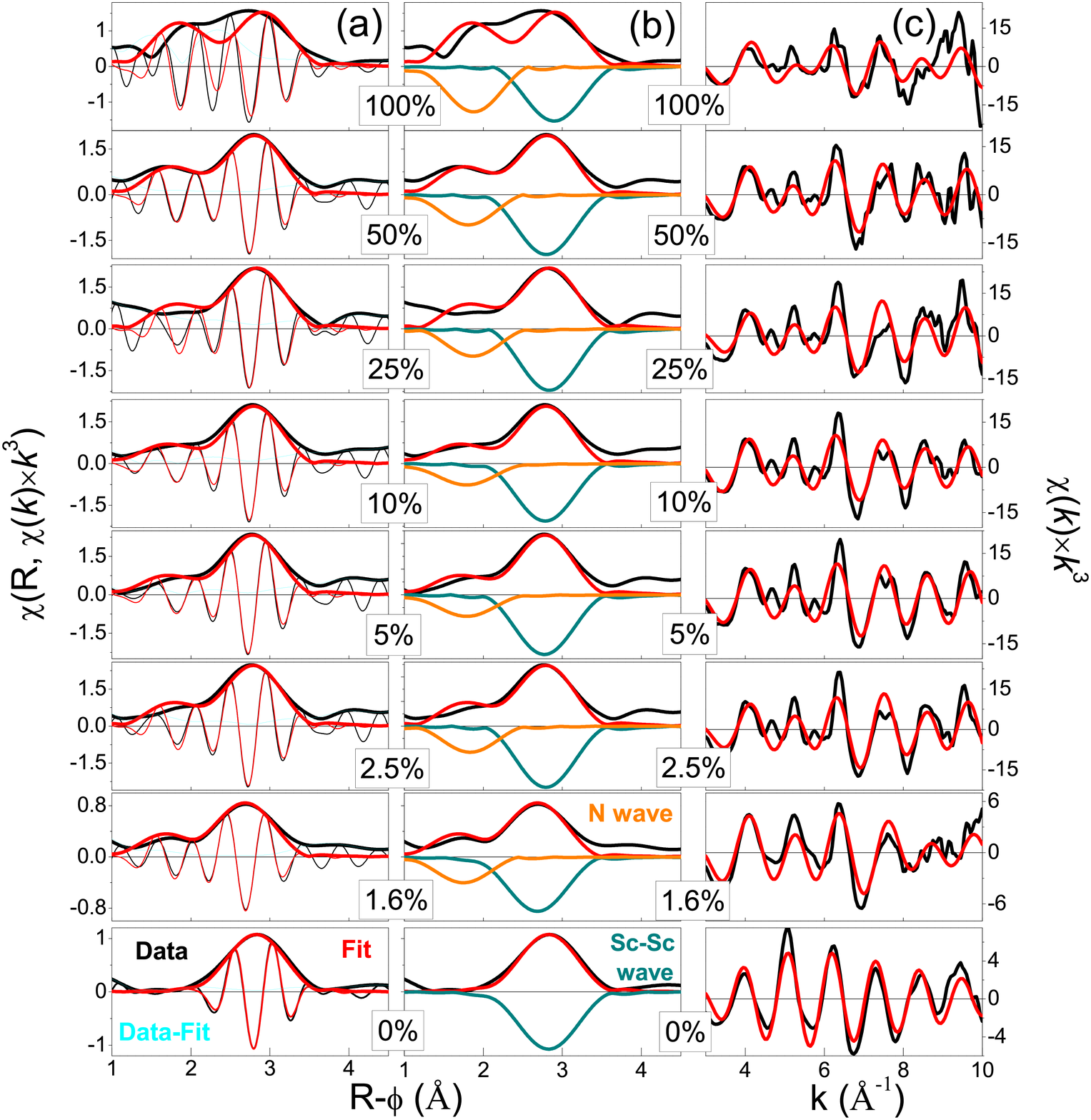}
	\vspace{-35mm}
	\caption{Comparison of Fourier transform (FT) moduli $ \chi $\,(\textit{R}) (a), Re [$ \chi $\,(\textit{R})] (b) in the \textit{R} range and $ \chi $(\textit{k})$ \times $\textit{k}$ ^{3} $ spectra in the \textit{k} range (c) of Sc and ScN thin film samples deposited at various \rn = 2.5, 5, 10, 25, 50 and 100{\p}.}
	\label{exafs}
\end{figure*}

\begin{table*}\center \caption{\label{EXAFSFIT} The metrical parameters obtained from the fitting of the EXAFS data recorded at Sc K-edge. Considering the central atom as Sc, here, N and N$ ^{'} $= first and second nearest neighbor co-ordination, R$ _\mathrm{Sc-N} $ and R$ _\mathrm{Sc-Sc} $ = atomic pair distance of the first and second neighbors i.e. Sc-N and Sc-Sc, $ \sigma $$ _\mathrm{Sc-N} $ and  $ \sigma $$ _\mathrm{Sc-Sc} $ = root mean square displacement obtained from fitting of the first and second shell.}
	\begin{tabular}{lllllll} \hline\hline
	{\rn} & N & R$ _\mathrm{Sc-N} $  & $ \sigma $$ _\mathrm{Sc-N} $ & N$ ^{\prime} $ & R$ _\mathrm{Sc-Sc} $ & $ \sigma $$ _\mathrm{Sc-Sc} $\\
		(\%) & & (\AA)  & (\AA) & & (\AA) & (\AA) \\\hline	
		0\%  & - & - & - & 7.23 & 3.25 & 0.098 \\
		  & - & - & - & ($ \pm $1.82) & ($ \pm $0.02) & ($ \pm $0.01)  \\\hline
		1.6\%  & 2.252 & 2.189 & 0.064 & 5.78 & 3.136 & 0.098 \\
		 & ($ \pm $0.676) & ($ \pm $0.025) & ($ \pm $0.031) & ($ \pm $1.59) & ($ \pm $0.021) & ($ \pm $0.017) \\\hline
		2.5\%  & 4.48 & 2.25 & 3.358 & 9.25 & 3.20 & 0.06 \\
		 & ($ \pm $1.34) & ($ \pm $0.02) & - & ($ \pm $2.55) & ($ \pm $0.02) & ($ \pm $0.02) \\\hline
		5\%  & 7.42 & 2.23  & 2.488 & 8.93 & 3.19 & 0.06 \\
		 & ($ \pm $2.22) & ($ \pm $0.03) & ($ \pm $0.03) & ($ \pm $2.42) & ($ \pm $0.02) & ($ \pm $0.02) \\\hline
		10\% & 8.03 &2.23 & 0.11 & 9.40 & 3.20 & 0.07 \\
		 & ($ \pm $2.4) & ($ \pm $0.03) & ($ \pm $0.03) & ($ \pm $2.6) & ($ \pm $0.02) & ($ \pm $0.02) \\\hline
		25\%  & 5.12 & 2.27 & 0.06 & 7.46 & 3.23 & 0.04 \\
		 & ($ \pm $1.5) & ($ \pm $0.03) & ($ \pm $0.03) & ($ \pm $2.14) & ($ \pm $0.02) & ($ \pm $0.02) \\\hline
		50\%  & 5.86 & 2.23 & 0.07 & 8.25 & 3.21 & 0.07 \\
		 & ($ \pm $1.75) & ($ \pm $0.03) & - & ($ \pm $2.3) & ($ \pm $0.02) & ($ \pm $0.02) \\\hline
		100\%  & 7.61 & 2.28 & 0.07 & 5.39 & 3.27 & 0.05 \\
		 & ($ \pm $2.28) & ($ \pm $0.02) &  & ($ \pm $1.62) & ($ \pm $0.02) & ($ \pm $0.03) \\
		\hline
	\end{tabular}
\end{table*}

Figure~\ref{exafs}(a) and (b) shows the Fourier Transform (FT) moduli $ | $$ \chi $(R)$ | $ and the real component [Re $\chi$(R)] of the Sc K-edge EXAFS spectra as a function of radial distance (R-$\phi$) and the corresponding best fit, whereas, Figure~\ref{exafs}(c) demonstrates the $ \chi $(\textit{k})$ \times $\textit{k}$ ^{3} $ spectra. For fitting, hcp and cubic rocksalt type NaCl structure of Sc and ScN were considered having space groups of P63/mmc~\cite{chowdhury2020study} and Fm$\bar{3}$m~\cite{casamento2019molecular}, respectively. The fitting was performed using LP obtained from the XRD data and the obtained metrical parameters are tabulated in Table~\ref{EXAFSFIT}. 

For Sc sample, the single shell of FT spectra corresponds to Sc co-ordinated to 7.2($ \pm $1.8) Sc atoms each having an atomic pair distance of 3.25($ \pm $0.02)\,{\AA}. Hence, the local co-ordination environment is rather in a distorted hexagonal symmetry which might be responsible for an intense pre-edge feature across the Sc K-edge absorption spectra (Figure~\ref{exafsnorm}). In addition, for ScN, considering a theoretical LP of a$ ^{\ast} $ = 4.501\,\AA~\cite{casamento2019molecular,deng2015optical}, the first and second nearest neighbor distances can be anticipated at R$ ^{^{\ast}} $$ _\mathrm{Sc-N} $ = a/2 = 2.25\,{\AA} and R$ ^{^{\ast}} $$ _\mathrm{Sc-Sc} $ = a/$ \sqrt{2} $ = 3.18\,{\AA}~\cite{tuilier2007structural}. As can be seen from Figure~\ref{exafs}(a) and (b), the two consecutive maxima distributed over R-$ \phi $ = 1~-~3.9\,{\AA} range correspond to the first Sc-N and second Sc-Sc nearest neighbor bonds. For 1.6{\p} sample, even though it was not possible to obtain the structural parameters from the XRD data due to appearance of a single broad peak, but the obtained EXAFS fitting parameters from Table~\ref{EXAFSFIT} clearly states the interstitial nature of the sample with less content of both N and Sc atoms in the first and second shell co-ordinations to fully evolve to the fcc phase considering only the ScN phase during fitting. Apparently, at higher {\rn} flow (above 2.5{\p}), all ScN samples typically exhibit the octahedral inversion symmetry as expected in NaCl structure and corroborates well with our XRD data within the experimental resolution, as expected. To take into account the optical and mechanical response of the ScN samples, UV-Vis measurements along with nanoindentation tests were performed and are discussed in section~\textbf{\ref{UVVIS}}.

\subsection{Optical \& Mechanical Behavior}\label{UVVIS}

\begin{figure*}\center
	\includegraphics
	[width=1\textwidth] {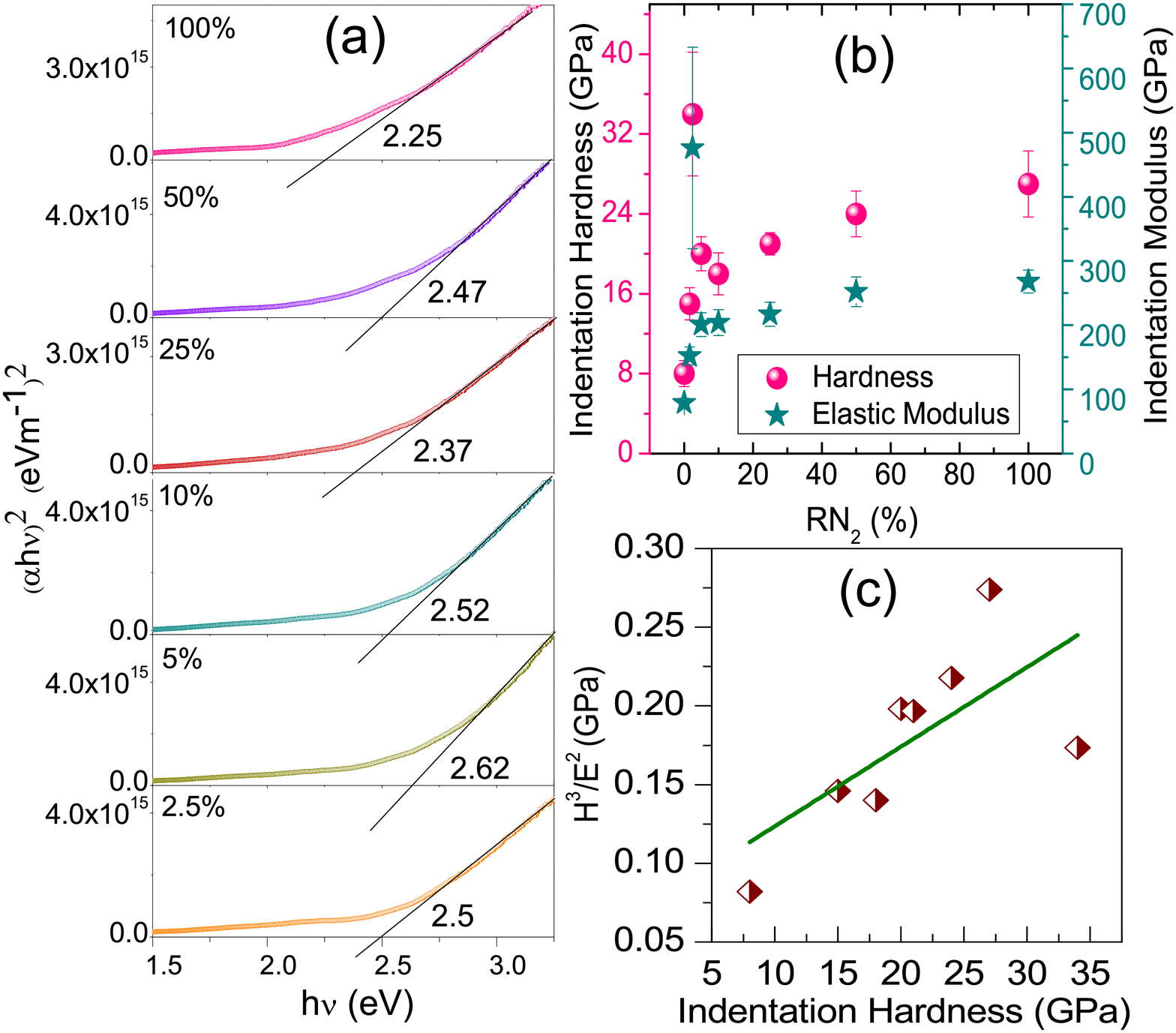}
	\vspace{-30mm}
	\caption{Obtained direct bandgaps from the Tauc's plot of the absorption co-efficient as a function of incident photon energy (a), nano-indentation hardness and modulus with error bars (b) and the ratio of (H$ ^{3} $/E$ ^{2} $) depicting resistance to plastic deformation of ScN thin film samples as a function of hardness (c), deposited at room-temperature at various \rn = 1.6, 2.5, 5, 10, 25, 50 and 100{\p} flow.}
	\label{bandgap}
\end{figure*}

In order to delve the optical properties with electronic structure, Figure~\ref{bandgap}(a) shows the Tauc's plot curve of $ \alpha $ as a function of incident photon energy for ScN thin film samples deposited at {\rn} = 2.5, 5, 10, 25, 50 and 100\,{\p} flows. The optical bandgap values were obtained  from the intersection of the energy axis by extrapolating the linear least square fitting curve around the inflection point using the Tauc relation, $ \alpha $h$ \nu $ = A (h$ \nu $ - E$ _\mathrm{g} $)$ {^{\frac{1}{2}}}$ for direct transitions~\cite{kuriyama1993optical} where, h$ \nu $ = photon energy, E$ _\mathrm{g} $ = optical bandgap and A is proportionality constant. As mentioned earlier in section~\textbf{\ref{SXAS}}, both Sc and 1.6{\p} ScN sample were metallic in nature and did not show any absorption in the whole energy spectrum. Beyond {\rn} = 1.6\,{\p}, all the samples showed semiconducting behavior with a well-defined optical bandgap. In this context, it is to be mentioned that ScN exhibits an indirect bandgap of 0.9\,eV~\cite{al2004surface,qteish2006exact}, whereas, two direct bandgaps at 2.2 and 3.8\,eV were reported~\cite{biswas2019development,deng2015optical}. Generally, only the first direct bandgap has been reported (see Table~\ref{techniques}) and will also be considered in this work. The large variation in the reported bandgap values have been attributed either to the formation of defect states near the conduction band for n-type degenerate semiconductor termed as `Burstein-Moss band filling effect'~\cite{burstein1954anomalous} or to the strain mediated effects which can modulate the energy band shifts to a certain extent~\cite{tamleh2018stress}. The trend was non-linear in nature with a maximum value of 2.62\,eV for 5{\p} ScN sample and a minimum of 2.25\,eV for sample deposited at {\rn} = 100{\p}. Thus, the variations in the bandgap values in the present study can be inferred primarily to the contribution of defects as the stress-strain mediated changes might be negligible in the present case, as all the samples were deposited on amorphous quartz substrates. It is worth mentioning here that the bandgap values of ScO$ _\mathrm{x} $N$ _\mathrm{y} $ and Sc$ _{2} $O$ _{3} $ are reported to be 3.25\,eV and 5.6\,eV, respectively~\cite{nayak2019rigid}, which are way higher than the obtained bandgaps in the present study.

Figure~\ref{bandgap}(b) demonstrates the measured indentation hardness (H) and modulus (E) of the ScN thin film samples as a function of {\rn} whereas, Figure~\ref{bandgap}(c) illustrates the ratio of (H$ ^{3} $/E$ ^{2} $) which corresponds to the resistance of ScN samples to plastic deformation as a function of H. As expected, Sc exhibits the lowest H and E values of 8($ \pm $1.3) and 79($ \pm $7)\,GPa, respectively. With nitridation, both H and E increases monotonically from 15~-~27\,GPa and 152~-~268\,GPa for the ScN samples, except for {\rn} = 2.5\%, which is close to the calculated value of 25\,GPa for ScN~\cite{aslam2018prediction}. At {\rn} = 2.5{\p}, the values maximizes at 34($ \pm $6.2) and 476($ \pm $157)\,GPa, respectively, which can be attributed to the highest density as estimated from the XRD data~\cite{caicedo2011mechanical}, smaller grain size and strong (111) and (222) texturing of the ScN sample~\cite{wu2012characteristics}. With increase in {\rn} = 2.5~-~50{\p}, from the XRD data (Figure~\ref{xrd}), it is evident that the XRD peaks shift chronically along the lower diffraction angle demonstrating in-plane compressive residual stress. In addition, the texturing effect is well retained, which is known to be the highest density plane and the corresponding H is considered to be the highest along the (111) plane~\cite{sarada2010highly} and are the reasons behind the increase in H. Even though, the texturing effect is witnessed for {\rn} = 100{\p} sample only along the (111) plane, but the decrease in grain size in turn elevates the grain boundaries which could lead to plausible rise in H  value~\cite{wu2012characteristics}. Furthermore, the gradual rise in E values from Sc$ \rightarrow $ScN can be attributed to the strong covalent bond formation of Sc-N than metallic Sc-Sc bonds in Sc~\cite{panda2019effects}, as evidenced from our XAFS study. Apart from this, the ScN thin film samples also show a propitious resistance to plastic deformation (H$ ^{3} $/E$ ^{2} $) following a linear trend with increase in H values~\cite{mayrhofer2003structure}. However, it is to be mentioned here that typically the values of H and E are likely comparable in the range of {\rn} = 2.5 - 100{\p} within the error bars.  

\section{Discussion}\label{DISCUSSION}

\begin{figure}\centering
 	\includegraphics
 	[width=0.3\textwidth] {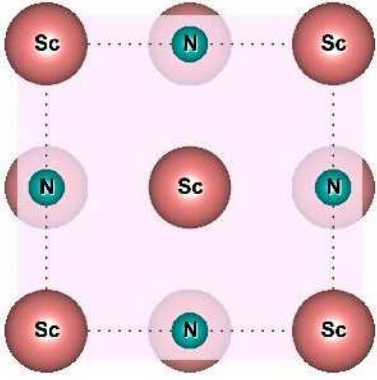}
 	\vspace{0mm}
 	\caption{Representative unit cell of ScN (100) crystal plane}
 	\label{plane}
 \end{figure}

At an early stage of subtle nitridation ({\rn} = 1.6{\p}), combined XRD and XAFS study reveal formation of an interstitial compound in the intermediate stage during evolution from hcp Sc to fcc ScN and with higher content of N, at {\rn} = 2.5{\p} and above, adaptation of fcc phase with octahedral symmetry were observed. Here, it is interesting to note that likewise in early TMNs, ScN does not possess bimetallic phases (e.g. Ti$ _{_{2}} $N, V$ _{2} $N and Cr$ _{_{2}} $N) at ambient temperature and pressure, and rather manifests a large homogeneity range (retaining NaCl type rocksalt crystal structure from as low as {\rn} = 2.5{\p} to as high as 100{\p}). In contrary, with increase in N$ _{2} $ atomic {\p}, the overall crystal lattice withstands a minimal lattice expansion of only $ \approx $\,1.7{\p} at highest {\rn} = 100{\p} for ScN. It could be well discerned in terms of higher interstitial lattice volume of ScN compared to other early TMNs in the series (TiN, VN and CrN). Since, the early TMNs exhibit NaCl type rocksalt crystal structure, the corresponding octahedral interstial site occupancy of nitrogen atoms would be at edge centers of the unit cell i.e. at ($ \frac{1}{2} $,0,0), (0,$ \frac{1}{2} $,0), (0,0,$ \frac{1}{2} $) and at the center i.e. ($ \frac{1}{2} $,$ \frac{1}{2} $,$ \frac{1}{2} $) position of the respective unit cell. Considering the (100) lattice plane of the unit cell, from the simple pictorial overview as shown in Figure~\ref{plane} for ScN sample, the radius of the interstitial site (R$ _\mathrm{int} $) can be similarly evaluated for early TMNs by solving two basic equations along the edge and diagonal as,
 
\begin{center}
	R$ _\mathrm{TM} $+2R$ _\mathrm{int} $+R$ _\mathrm{TM} $ = a,   .....(i)
\end{center}
and,
\begin{center}
	R$ _\mathrm{TM} $+2R$ _\mathrm{TM} $+R$ _\mathrm{TM} $ = (a$ ^{2} $ + a $^{2} $)$ ^{\frac{1}{2}} $,   .....(ii)
\end{center}

where, R$ _\mathrm{TM} $ = radius of TM atom, and a = LP of TMN. Hence, the corresponding structural parameters of the early TMNs are enlisted in Table~\ref{interstitial}.

\begin{table*} \centering \caption{\label{interstitial} Early transition metals (TM) and their corresponding crystal structures (CS) and lattice parameters (LP$ _\mathrm{TM} $). In comparison to metal counterparts, lattice parameters (LP$ _\mathrm{TMN} $) of their nitrides with calculated radius of interstitial octahedral site (R$  _\mathrm{int} $)  have been tabulated below.}
	\begin{tabular} {lllllll} \hline
		TM & CS & LP$ _\mathrm{TM} $ & TMN & LP$ _\mathrm{TMN} $ & R$  _\mathrm{int} $  & Ref.\\
		& & ({\AA}) & & ({\AA}) & ({\AA})  & \\\hline
		Sc & hcp & a = b = 3.309, & ScN & 4.501  & 0.659  & \cite{al2000molecular} \\
		&  & c = 5.273 & & &  & \\
		Ti & hcp & a = b = 2.951, & TiN & 4.24  & 0.621  & \cite{patsalas2000effect} \\
		&  & c = 4.684 & & &  & \\
		V & bcc & a = 3.03 & VN & 4.139 & 0.606  & \cite{caicedo2011mechanical} \\
		Cr & bcc & a = 2.885 & CrN & 4.14 & 0.606  & \cite{olaya2006influence}\\\hline
	\end{tabular}
\end{table*}

From Table~\ref{interstitial}, it is evident that ScN exhibits the largest unit cell with higher fraction of interstitial volume among the early TMNs, leading to a large homogeneity range for retainig the fcc rocksalt phase accompanied with minuscule lattice expansion. In this context, it is worth mentioning that Al et al. have reported that ScN can withstand upto $\approx$ 20{\p} of N vacancies within the crystal lattice~\cite{al2002phase}. Such a large homogeneity range has also been witnessed for other early TMNs like TiN$ _{x} $ (0.67 $ \leq $ \textit{x} $ \leq $ 1.3)~\cite{schramm2016impact}, VN$ _{x} $ (0.79 $ \leq $ \textit{\textit{x}} $ \leq $ 0.96)~\cite{pompe1982some} etc. in the octahedral symmetry. In addition, combined SXAS study at Sc L-edge, N K-edge and O K-edge reveal that the effect of oxidation is less pronounced for nitrided samples compared to their metal/interstitial counterparts. This is due to formation of strong Sc-N covalent bonds which results in relatively high Gibb's energy for oxide formation (-6.48\,eV) of ScN than metallic Sc-Sc bonds in pure Sc (-9.43\,eV) at $ \approx $298\,K~\cite{more2020correlation}. Furthermore, our Sc K-edge XANES spectra confirms distinct evolution from hexagonal to octahedral symmetry complemented with a clear rise in the valence state for ScN samples consistent with our XRD results. The pre-edge features of neither Sc nor ScN resembles the Sc K-edge oxide spectra~\cite{chasse2018influence} emphasizing on the higher surface oxidation effect in all these samples. Nonetheless, the local electronic structure as recorded from EXAFS replicates the XRD data with further insight on the presence of local defects in the vicinity of Sc atoms but the visible changes are however marginal for all the samples. Even so, the non-monotonic variations in the optical bandgaps across the whole {\rn} range have no clear trend and in this scenario, it is difficult to correlate them in terms of defects as in case of polycrystalline thin film samples, the role of defects are always expected to be higher than epitaxial ScN thin film samples studied so far. Howbeit, the bandgap values in the present study lie well within the energy regime of ScN thin film samples as observed for high {\Ts} depositions on single crystal substrates. Additionally, the hardness and indentation modulus agrees well during the evolution from Sc$ \rightarrow $ScN with a linearly increasing trend of resistance to plastic deformation and also matches well with those reported in the literature~\cite{gall1998growth,moram2006young} within the experimental error bars.  

\section*{Conclusion}

In lieu of adoptation of high tempearture depositions hitherto, as-deposited ScN thin film samples exhibited highly textured orientation along the (111) and (222) reflection planes grown on amorphous quartz substrates due to preferable highest surface energy configuration at low temperature regime (here 300\,K). SXAS study reveals pronounced incorporation of oxygen in metallic Sc and interstial ScN sample deposited at {\rn} = 1.6{\p} flow, although reduction of oxygen content can be witnessed with nitridation of the samples ({\rn} = 2.5~-~100{\p}). Complementary XAFS study shows distinct evolution from Sc$ \rightarrow $ScN, where an intense pre-edge feature stems from non-centrosymmetric distortion for Sc and interstitial ScN ({\rn} = 1.6{\p}) sample, whereas, octahedral symmetry was retained by rest of the ScN samples deposited at higher {\rn} flow. From UV-Vis measurement, the obtained direct optical bandgaps were found to vary between 2.25 - 2.62\,eV for {\rn} = 2.5~-~100{\p}, well in agreement with the values routinely reported in literatures for epitaxial ScN thin film samples. Even so, the nano-indentation measurements validates the high hardness of highly elastic ScN thin film samples ranging between 15~-~34\,GPa with a monotonically increasing trend in the value of resistance to plastic deformations. In turn, the large homogenity range of Sc-N system has been compared with elemental early 3d transition metal nitride series viz. TiN, VN and CrN to comprehend the phase stability of cubic NaCl rocksalt type structure of ScN thin film samples over large variation in {\rn} flow. Hence, cumulative inferences drawn from this work can be epitomized as high vacuum deposition is imperative for high quality ScN samples but an alternate room temperature deposition can be adopted as opposed high {\Ts }, to minimize the unintentional diffusion mechanisms at elevated temperatures finding applications in electronics viz. CMOS integrated circuits, free standing films on plastic substrates. 
     
\section*{Acknowledgments}
Authors (SC and RG) are grateful to UGC-DAE CSR, Indore for
providing financial support through
CSR-IC-BL-62/CSR179-2016-17/843 project. Thanks are due to V. R.
Reddy and Anil Gome for XRR measurements, Rakesh Sah for
SXAS measurement at BL-01, Indus 2, RRCAT and Layanta Behera for various experiments. SC is thankful to Sanjay Nayak for fruitful discussions and Yogesh Kumar for his extended help in XRD and XAFS measurements. We also thank S. Tokekar, A. J. Pal, A. K. Sinha, D. M. Phase, V. Ganesan and V. Sathe for their kind
support and constant encouragements.


\end{document}